# A New Atomistic Force Field for Pyridinium-Based Ionic Liquids: Reliable Transport Properties.

**Iuliia V. Voroshylova[1] and Vitaly V. Chaban[2]**

[1]CIQ/REQUIMTE – Department of Chemistry and Biochemistry, Faculty of Sciences, University of Porto, Rua do Campo Alegre 687, 4169-007, Porto, Portugal

[2]MEMPHYS – Center for Biomembrane Physics, Syddansk Universitet, Odense M., 5320, Kingdom of Denmark

**Abstract**. Reliable force field (FF) is a central issue in successful prediction of physical chemical properties via computer simulations. This work introduces new FF parameters for six popular ionic liquids (ILs) of the pyridinium family (butylpyridinium tetrafluoroborate, bis(trifluoromethanesulfonyl)imide, dicyanamide, hexafluorophosphate, triflate, chloride). We elaborate a systematic procedure, which allows to account for specific cation-anion interactions in the liquid phase. Once these interactions are described accurately, all experimentally determined transport properties can be reproduced. We prove that three parameters per interaction site (atom diameter, depth of potential well, point electrostatic charge) provide a sufficient basis to predict thermodynamics (heat of vaporization, density), structure (radial distributions), and transport (diffusion, viscosity, conductivity) of ILs at room conditions and elevated temperature. The developed atomistic models provide a systematic refinement upon the well-known Lopes—Padua (CL&P) FF. Together with the original CL&P parameters the present models foster a computational investigation of ionic liquids.

**Introduction**

Atomistic-precision simulations constitute a relatively large portion of published reports in the vibrant research field of ionic liquids (ILs).[1-8] A basic search performed in the ISI Web of Knowledge database on May 31, 2014 reveals that 1,863 out of 35,252 papers devoted to ILs report computer simulation studies. The papers indexed during the last five years, 2009 to 2014, were included in this statistical investigation. One can conclude that simulations have become a solid supplement to the experimental studies, providing key physical insights in the complicated condensed-phase phenomena. Reliable sampling methodology and force field (FF) are a central issue in successful prediction of physical chemical properties via various sorts of computer simulation techniques.

Several developments for the state-of-the-art atomistic description of ionic liquids have been introduced over the recent years. Canongia Lopes and Padua were essentially pioneers in elaborating a comprehensive systematical methodology and a force field for ionic liquids (CL&P). A large variety of IL families have been parametrized thus far within the framework of this methodology.[5,7,9-13] The developed models are internally consistent, transferrable, and compatible providing a good accuracy of the most simulated thermodynamic and structure properties at room conditions and somewhat elevated temperatures. Another comprehensive contribution to IL simulations has been delivered by Sambasivarao and Acevedo.[14] Sixty eight unique combinations of room temperature ILs featuring 1-alkyl-3-methylimidazolium, N-alkylpyridinium, choline cations, supplemented by an extensive set of anions, have been parametrized using harmonic bonded forces and additive pairwise potential functions. The developed models have been tested using the Metropolis Monte Carlo simulations. In turn, Borodin has applied a Drude oscillator model to produce polarizable ions for a significant number of ILs.[15] A set of simulated thermodynamic (density, heat of vaporization) and transport properties (self-diffusion coefficients, shear viscosity, ionic conductivity) appear in a reasonable agreement with the available experimental results. We have previously made certain contribution to the FF development and refinement by coupling electronic structure and molecular mechanical descriptions of inter-ionic interactions.[2,16-18]

The present work further elaborates an approach to develop efficient atomistic force field models for ILs under a simple approximation of additive non-bonded interactions. Starting from the FF for isolated ions in vacuum developed by Canongia Lopes and Padua (CL&P),[5,7] we implement (1) a correction for electron transfer effects between the cation and the anion and (2) a correction for cation-anion atom-atom contact distances. Both corrections are driven by the results of hybrid density functional theory (DFT). We apply our methodology to six ILs representing pyridinium family: (1) butylpyridinium tetrafluoroborate [C$_4$PY][BF$_4$]; (2) butylpyridinium bis(trifluoromethanesulfonyl)imide [C$_4$PY][TFSI]; (3) butylpyridinium dicyanamide [C$_4$PY][DCA]; (4) butylpyridinium hexafluorophosphate [C$_4$PY][PF$_6$]; (5) butylpyridinium triflate [C$_4$PY][TF]; (6) butylpyridinium chloride [C$_4$PY][Cl]. Figure 1 introduces the ions, which were used to compose the considered ILs. Certain experimental data for four of these ILs are available. These data can be used to evaluate the accuracy of the derived FF. We have not found any experimental properties for [C$_4$PY][DCA] and [C$_4$PY][PF$_6$]. As our simulations indicate, [C$_4$PY][DCA] represents a peculiar case with low viscosity and high ionic conductivity. Experimental efforts to address transport properties of this IL are welcome.

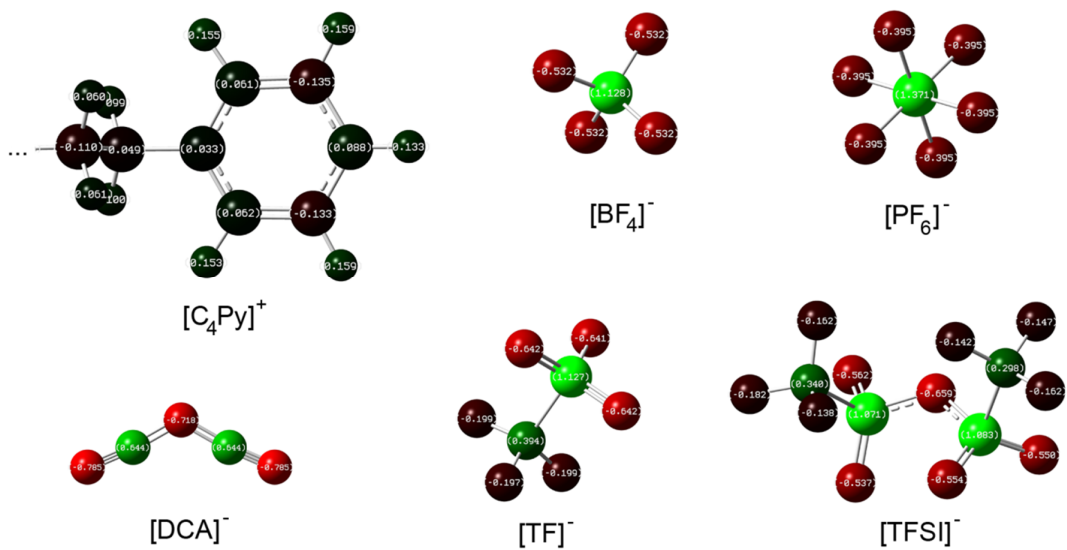

Figure 1. Optimized geometries of ions, which were used to compose ionic liquids investigated in this work: butylpyridinium, tetrafluoroborate, hexafluorophosphate, dicyanamide, triflate, bis(trifluoromethanesulfonyl)imide. The atoms are colored with respect to their electrostatic charge. Red atoms are electron-rich, green atoms are electron-poor, and dark colors correspond to electrostatically neutral parts of the molecule.

**Computational methodology**

The thermodynamic (heat of vaporization, $\Delta H_{vap}$, mass density, $d$), structure (radial distribution functions), and transport properties (self-diffusion coefficients, $D_{\pm}$, shear viscosity, $\eta$, ionic conductivity, $\sigma$) have been obtained using conventional molecular dynamics (MD) simulations. The basic list of the simulated systems is provided in Table 1. The IL ions were placed in cubic periodic MD boxes, whose densities were calculated to correspond to ambient pressure at the requested temperature (298, 313, 338, 353, 403 K). The simulated temperatures were selected in view of the currently available experimental data and the aggregation state of particular IL at given temperature. For instance, transport properties of butylpyridinium chloride were simulated at 353 and 403 K, because this IL is in the glassy (or even solid) state around 298-338 K. Only the heat of vaporization is presently available in literature for [C$_4$PY][Cl]. Ionic conductivities, representing a collective ionic property, require an extensive sampling to obtain results of reasonable precision. For this reason, conductivities were derived at two temperatures only. An interested reader can smoothly extend the investigation to any temperature or pressure of interest using freely available parameters. In turn, shear viscosities are more abundant in literature and can be measured using a relatively simple apparatus. Therefore, we provide viscosities for all developed models at five temperatures (298-403 K) to facilitate a straightforward comparison.

Each system at each temperature was sampled during 80 ns. The first 10 ns were regarded as equilibration, whereas consequent 70 ns were used to derive properties. Cartesian coordinates were saved every 5 ps and pressure tensor components (for viscosity calculation) were saved every 0.02 ps. More frequent saving of trajectory components was preliminarily tested, but no systematic accuracy improvement was found. Self-diffusion coefficients were computed from mean-square displacements of atomic positions and conductivity was computed from mean-square displacements of translational dipole moment (the Einstein-Helfand fit). Prior to these calculations, the trajectories were pre-processed to remove information about periodic boundary conditions. Additionally, viscosity was calculated using cosine-shape acceleration of all ions. The mathematical foundation of this method and its applications to molecular dynamics simulations were in detail discussed by Hess.[19] The first nanosecond of the simulation was used for an accelerated ionic flow to be established. The subsequent 19 ns were used for the viscosity calculation.

Table 1. The list of systems simulated in the present work. The quantity of ion pairs per IL was selected with respect to the cation and anion sizes

| # | IL | Number of ion pairs | Number of interaction centers | Experimental Data Available |
|---|---|---|---|---|
| 1-6 | [C$_4$PY][TFSI] | 125 | 4875 | $\Delta H_{vap}$, $d$, $D_{\pm}$, $\eta$, $\sigma$ |
| 7-12 | [C$_4$PY][BF$_4$] | 150 | 4350 | $\Delta H_{vap}$, $d$, $D_{\pm}$, $\eta$, $\sigma$ |
| 13-18 | [C$_4$PY][DCA] | 150 | 4350 | none |
| 19-24 | [C$_4$PY][PF$_6$] | 125 | 3875 | none |
| 25-30 | [C$_4$PY][TF] | 125 | 4000 | $d$, $\eta$, $\sigma$. |
| 31-33 | [C$_4$PY][Cl] | 175 | 4375 | $\Delta H_{vap}$ |

All systems were simulated in constant-pressure constant-temperature ensemble. The equations of motion were propagated with a time-step of 2 fs. Such a relatively large time-step was possible due to constraints imposed on the carbon-hydrogen covalent bonds (instead of a harmonic potential). The electrostatic interactions were simulated using direct Coulomb law up to 1.3 nm of separation between the interaction sites. The electrostatic interactions beyond 1.3 nm were accounted for by computationally efficient Particle-Mesh-Ewald (PME) method. It is important to use PME method in case of ionic systems, since electrostatic energy beyond the cut-off usually contributes 40-60% of total electrostatic energy. We used a smaller real-space cut-off, as compared to the original work of Canongia Lopes—Padua. This computational trick does not change the accuracy significantly, according to our preliminary assessment, but allows to speed up the MD calculations. The Lennard-Jones-12-6 interactions were smoothly brought down to zero from 1.1 to 1.2 nm using the classical shifted force technique. The constant temperature (298, 313, 338, 353, 403 K) was maintained by the Bussi-Donadio-Parrinello velocity rescaling thermostat[20] (with a time constant of 0.5 ps), which provides a correct velocity distribution for a statistical mechanical ensemble. The constant pressure was maintained by Parrinello-Rahman barostat[21] with a time constant of 4.0 ps and a compressibility constant of $4.5 \times 10^{-5}$ bar$^{-1}$. All molecular dynamics trajectories were propagated using the GROMACS simulation suite.[22] Analysis of thermodynamics, structure, and transport properties was done using the supplementary utilities distributed with the GROMACS package,[22] where possible, and the in-home tools.

Cation-anion specific interactions were unveiled using the hybrid DFT functional. We describe an electron density using the recently proposed high-quality hybrid exchange-correlation functional, omega B97X-D.[23] The exchange energy is combined with the exact energy from the Hartree-Fock theory. Empirical atom-atom dispersion correction is included. According to Chai and Head-Gordon,[23] the omega B97X-D functional simultaneously yields an improved accuracy for thermochemistry, electron kinetics, and non-covalent interactions. We construct the wave function using a comprehensive Dunning's correlation-consistent basis set. Based on our preliminary tests, triple-zeta basis set from this family, augmented with the diffuse functions (aug-cc-pVTZ), provides an accurate description of the electronic density and energy levels. Certain known defects, such as excessive valence electron delocalization, widely observed in case of pure DFT methods, were not reported for the presently selected method and basis set. Electrostatic potential (ESP), derived from the electronic wave function, was fitted using a set of point charges localized on each atom of the cation and anion, including hydrogen atoms. In case of ion pairs (i.e. neutral systems), an additional constraint was imposed during the fitting procedure to rigorously reproduce dipole moment of the ion pair. Atomic radii were assigned on the basis of the CHELPG scheme.[24]

**Performance of the original force field**

Table 2 provides a summarized comparison of the thermodynamic and transport properties derived using the CL&P FF vs. the available experimental data. The simulated thermodynamic properties exhibit satisfactory to very good agreement with the experimental values. Unfortunately, transport properties are not reproduced well. The diffusion coefficients and ionic conductivities are underestimated by up to five times. Consequently, the viscosities are proportionally overestimated. Note, that comparison has been made at somewhat elevated temperature, 353 K. With a temperature decrease, the transport property prediction errors will increase drastically. The physical reason of this observation and trend is clear. The force fields developed for isolated cation and anion in vacuum do not account for electronic polarization and charge transfer effects, which play a significant role for any ionic system in condensed state. In case of ionic liquid, these effects are particularly important, because valence electron density is well delocalized over both the cation and the anion (Figure 2). This electronic orbital delocalization can be also interpreted as certain degree of covalence in the cation-anion bonding. In most cases, the neglect results in the unnaturally elevated pairwise interaction energies and higher activation barriers than in real systems. Thermodynamic properties and most structure correlations are moderately susceptible to somewhat stronger non-bonded interactions. However, transport properties can alter by as much as orders of magnitude following just a 10-20% increase of vaporization heat or a conjugated energetic property. It also depends on the structure of particular ion and how close the simulated temperature is located to glass transition temperature of the corresponding compound. The goal of the present work is to introduce implicit correction factors, so that to attend the charge transfer effects and improve the accuracy of the simulated transport properties.

Table 2. Thermodynamic and transport properties at 353 K derived using the CL&P force field. The available experimental properties are shown in parentheses for the sake of comparison

| IL | $d$, kg m$^{-3}$ | $\Delta H_{vap}$, kJ mol$^{-1}$ | $D_+$, $10^{11}$ m$^2$s$^{-1}$ | $D_-$, $10^{11}$ m$^2$s$^{-1}$ | $\eta$, $10^1$ cP | $\sigma$, S m$^{-1}$ |
|---|---|---|---|---|---|---|
| [C$_4$PY][TFSI] | 1442 (1403) | 167 (138) | 2.3 (14) | 2.0 (11) | 5 (1) | 0.3 (1.6) |
| [C$_4$PY][BF$_4$] | 1139 (1176) | 162 (167) | 1.7 (1.9$^{313K}$) | 1.0 (2.1$^{313K}$) | 5 (1.58) | 0.5 (1.3$^{338K}$) |
| [C$_4$PY][DCA] | 1024 | 162 | 2.8 | 3.0 | 4 | 0.8 |
| [C$_4$PY][PF$_6$] | 1300 | 174 | 0.7 | 0.3 | 40 | 0.1 |
| [C$_4$PY][TF]* | 1293 (1284) | 163 | 0.7 | 0.4 | 10 (2.5) | 0.5 (0.98) |
| [C$_4$PY][Cl] | 1009 | 189 (170) | 0.2 | 0.1 | 50 | 0.1 |

* The properties of [C$_4$PY][TF] were derived at 338 K to make a point-to-point comparison at this temperature.

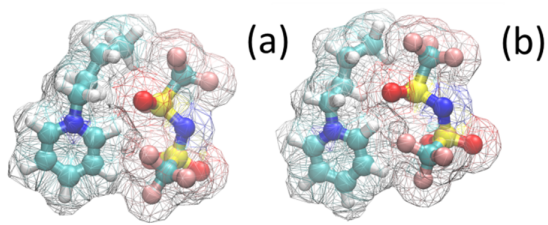

Figure 2. (a) Delocalization of the highest occupied molecular orbital (HOMO) and (b) the lowest unoccupied molecular orbital (LUMO) of [C$_4$PY][TFSI] ion pair. Both orbitals are shared by the [C$_4$PY]$^+$ cation and [TFSI]$^-$ anion, indicating their strong coupling upon energetically favorable orientation.

**Force field development**

The force field development was based on the following philosophy. First, we start with the CL&P FF, since it features the most general and unbiased derivation procedure out of all presently available models. Second, minimum possible number of refinements per IL should be made. Third, no empirical data should be used during FF derivation (although certain experimental data are available, Table 2). This condition is required to preserve generality of the proposed procedure. Fourth, the derivation procedure should be equivalent, as much as possible, for all ILs. However, parameters of butylpyridinium cation can be different when combined with different anions, because many anions exhibit significant impact on the electronic structure of the cation. Fifth, only three mathematically independent parameters are allowed per each pairwise interaction to describe the system. All pairwise interactions are obtained using the Lorenz-

Berthelot combination rules and the Coulomb law. No many-body interactions are allowed. Sixth, a set of selected properties, representing thermodynamic, structure and dynamic properties, is used to conclude whether the simulated IL is a proper approximation of the natural IL. If the force field reproduces heat of vaporization, density, key radial distribution functions, self-diffusion coefficients, shear viscosity, and ionic conductivity over the temperature range of interest, it is assumed to reproduce also other properties, which were not addressed during the FF validation.

The electronic structure of the pyridinium cation is truly unique. It can scarcely be predicted on the basis of conventional chemical wisdom. The ring is composed of five carbon atoms, whereby the first and the fifth atoms are linked by a single nitrogen atom. It is often assumed that a positive charge mainly comes from this nitrogen atom. However, electronic structure quantum chemical description indicates that most electron deficient atoms are carbon atoms linked to nitrogen. This effect can be quantified, for instance, in terms of the Mulliken charges computed from the converged wave function for an isolated cation. The C-H groups bonded to nitrogen exhibit +0.42e and +0.41e charges, whereas nitrogen exhibits +0.01e, being fairly neutral. Additionally, α-carbon of the alkyl chain is electron deficient, +0.21e, whereas β–carbon is almost neutral, +0.04e. The point charges, obtained from fitting of the ab initio electrostatic potential, exhibit a similar trend, with nitrogen atom, again, being neutral, +0.03e. The charges of neighboring carbon atoms are +0.06e, while their hydrogen atoms are somewhat more positive, +0.15e. Note, that all these charges are very small, as compared to the ESP charges in traditional polar compounds. For instance, ESP charges obtained using the same electronic structure method for water molecule in polarizable medium are -0.8e (oxygen) and +0.4e (hydrogen). Inorganic metal ions possess ±1,2,3, depending on their valence.

Similar small charges of carbon, hydrogen, and nitrogen atoms of pyridinium ring are due to valence electron delocalization. The described electron configuration is highly susceptible to the presence of external electric field. The point charges and higher electron energy levels shift significantly depending on the surrounding ions or polar molecules. Such an electron density distribution makes it challenging to develop fixed-charge force fields for molecular mechanics simulations. A radial workaround can be to avoid Coulombic part of the potential, transferring its role to the short-range interactions. This methodological possibility, however, must be extensively validated. The CL&P FF proposes a positive nitrogen atom (+0.15e) with nearly neutral carbon atoms (-0.07e, +0.02e, 0.00e) and all positive hydrogen atoms (+0.15e). This charge assignment is not supported by our ab initio results, but provides a reasonably accurate IL structure. We stick to the suggested point charge distribution, but refine it in accordance with the anion impact for each IL separately. Figures S1-S2 demonstrate different response of the $[C_2PY]^+$ to different anions, as follows from omega B97XD hybrid DFT description. An additional refinement concerns non-bonded interatomic distances between certain atoms of cation and anion, where these distances cannot be adequately described by originally proposed non-bonded parameters. The following section briefly describes introduced refinements, as applied to various cation-anion pairs.

*Butylpyridinium bis(trifluoromethanesulfonyl)imide*. As a result of the cation-anion electrostatic interaction, point charges of anion atoms decrease. The symmetry of electron density is expectedly broken. For instance, oxygen atoms of $[TFSI]^-$ gain -0.51e, -0.47e, -0.39e, -0.50e, while they all have -0.55e in isolated anion. In turn, the charge of nitrogen decreases from -0.66e (isolated state) to -0.55e (ion pair state). Interestingly, the charge of C(F3) increases from +0.31e to +0.36e, while fluorine charge decreased from -0.15e to ca. -0.13e. Sulfur, as the only third-period element in its environment is expected to be more polarizable. Indeed, it decreases its charge from +1.09e to +0.92e and +0.74e (in the moiety coordinating cation). The higher charge transfer potential of sulfur must be accounted for in the refinement procedure by introducing a separate charge scaling factor. No atom of the anion alters the sign of its charge due to formation of ion pair. However, the charge sign alteration is observed in the case of pyridinium cation. Two carbon atoms acquire electron density becoming -0.18e and -0.17e, whereas the other three atoms are positive: +0.08e, +0.07e, and +0.02e. The nitrogen atom becomes four times more positive, +0.13e, since it participates in the coordination (Figure 1).

The sum of the ESP charges of the anion atoms obtained in this way is -0.85e. In turn, ESP charges of the pyridinium ring, α–carbon and β–carbon atoms of the butyl chain were uniformly scaled to provide +0.85e in total. The neutral atoms of the hydrocarbon chain were not modified, as the chain does not participate in the anion coordination.

Furthermore, we compare non-bonded cation-anion distances, as provided by CL&P FF for an ion pair, with the same distance from the geometrically optimized structure using hybrid density functional theory. Both structures correspond to internal energy minimum at zero temperature, so the presented comparison is direct. The distance predicted by the CL&P FF is slightly overestimated, since the CL&P FF does not account for an electron transfer effect. Compare, the shortest nitrogen (cation)—oxygen (anion) distances equal to 3.21 and 3.31 Å (CL&P), while hybrid DFT functional suggests 3.07 and 3.14 Å. Similarly, hydrogen(α-C)—oxygen(anion) distance is 2.62 Å (CL&P) vs. 2.34 Å (DFT). We do not currently modify non-bonded parameters responsible for these distances, since we assume that this average deviation can be implicitly corrected after an assignment of the refined point electrostatic charges. However, this refinement can be done following our FF derivation philosophy.

*Butylpyridinium tetrafluoroborate*. [BF$_4$]$^-$ anion is equipped with highly polar B-F covalent bonds resulting in +1.13e and -0.53e ESP charges on boron and fluorine atoms, respectively. This anion strongly influences the pyridinium ring coordinating the first carbon atom of the butyl chain. The charges of boron and fluorine atoms decrease significantly, providing +0.84e and -0.42e. We observed this same effect in the case of imidazolium-based cations.[2,16] The alteration of charges localized at the cation atoms is small by absolute value and can be understood from the geometry of the ion pair (Figure 3). Similar to [TFSI]$^-$, we have uniformly decreased charges of pyridinium ring to obtain -0.84e. Compare, the scaling factor of 0.84 is very close to the scaling factors that we discussed in the case of 1-ethyl-3-methylimidazolium and 1-butyl-3-methylimidazolium tetrafluoroborate ILs.[16]

[BF$_4$]$^-$ exhibits quite a defined contact with the cation, fluorine—hydrogen(pyridinium ring). The corresponding distances, obtained from the hybrid DFT functional, are 2.27 and 2.31 Å, while the lengths are overestimated by the original FF, 2.57 and 2.67 Å. The H-F cross-sigma parameter was decreased accordingly to allow closer approach. Note, that the radius of hydrogen atom, due to its unique structure and nature of bonding, is poorly defined. Its alteration in various strong interactions can be expected.

*Butylpyridinium dicyanimide.* [DCA]$^-$ is coordinated above nitrogen and its neighboring carbon atoms of the pyridinium ring. Interestingly, [DCA]$^-$ attains to the symmetry of –C-N-C- atoms in the cation structure. The ESP charges slightly decrease indicating a modest electron density transfer. Compare, -0.67e vs. -0.72e (central nitrogen atom), -0.70e vs. -0.79e (side nitrogen atoms). The charge on the carbon atoms remains unchanged and retains the original symmetry, +0.64e. The cation-anion distances are modified insignificantly. Nitrogen(anion)—hydrogen(ring) distance is 2.66 Å (CL&P) and 2.51Å (DFT). The corresponding non-bonded interaction was not refined, as it unlikely plays a principal role in the condensed phase structure of this IL.

*Butylpyridinium hexafluorophosphate.* With a highly electron deficient phosphorus atom (+1.37e), [PF$_6$]$^-$ to a larger extent resembles classical inorganic anions than those giving rise to ionic liquids. However, [PF$_6$]$^-$ containing ILs possess an extended liquid state range, though not all of them are liquid at room conditions. Pairing with the pyridinium cation decreases ESP charges to +1.13e (P) and -0.33e (F), while nitrogen acquires +0.12e and carbon atoms bonded to it acquire -0.14e and -0.12e. All charges of both the anion and the pyridinium ring were scaled proportionally. There is a critical disagreement between fluorine(anion)-hydrogen(cation ring) distances in CL&P (2.87 Å) and DFT (2.09 Å). A similar large difference is observed in case of the fluorine(anion)—hydrogen(α-carbon) contact: 2.69 Å (CL&P) vs. 2.28 Å (DFT). The distances from the DFT calculation suggest an existence of hydrogen bonding in this pair of atoms. The corresponding cross-sigma parameter was decreased to reproduce the length of this hydrogen bond more realistically. The present refinement is seen highly important to provide a reasonably density of [C$_4$PY][PF$_6$]. Unfortunately, the experimental data for this interesting compound are yet missed.

*Butylpyridinium triflate*. The major considerations concerning [TFSI]$^-$ apply qualitatively to this case. The ESP charge of sulfur decreases from +1.13e to +1.01e due to ion pair formation. In turn, the ESP charge of oxygen decreases from -0.64e to -0.56e. Fluorine atoms acquire modest, but negative charge, -0.20e (isolated anion) and -0.15e (ion pair). No specific anion-cation coordination, such as in the case of [PF$_6$]$^-$, is found. If more than one ion pair is considered, the effect from ion pairing will be obviously somewhat different. However, qualitative features are well captured by the present description. The cation charges (excluding neutral hydrocarbon chain) were obtained through scaling by 0.78 to preserve system neutrality.

*Butylpyridinium chloride*. [Cl]$^-$ forms a strong hydrogen bond with the hydrogen atom of the pyridinium ring. The length of this bond at zero-temperature equals to 2.10 Å. It cannot be reproduced by the CL&P FF, which predicts a much higher length, 3.52 Å. That is, the separately parametrized cation and anion cannot form a strong hydrogen bond when mixed within the same system. It is, in our opinion, a very significant feature, which must be taken into account in MD simulations of ILs. The refinement procedure for butylpyridinium chloride IL cannot be based on the uniform scaling of ESP charges of the pyridinium ring. Instead, we adjusted an ESP charge (+0.01e) on one hydrogen atom of the ring, while a symmetric hydrogen atom was left with +0.15e charge (CL&P FF). This setup allows to avoid competition between two physically identical hydrogen atoms, which results in the location of [Cl]$^-$ above the pyridinium ring (CL&P). The new charge of [Cl]$^-$ was set according to the ESP charge in the 2×[C$_2$PY][Cl] system. The optimized geometry of this system is depicted in Figure S3.

Noteworthy, two ([PF$_6$]$^-$, [Cl]$^-$) of six investigated anions exhibit a very well-defined coordination site with respect to the pyridinium cation. [BF$_4$]$^-$, with the H–F hydrogen bond length of 2.27 Å, is in the intermediate position, while [DCA]$^-$, [TF]$^-$ and [TFSI]$^-$ are coordinated rather by a ring than by a particular atom. The absence of clear coordination site is probably a key feature making an ionic compound an ionic liquid. In the case of inorganic anions, hydrogen atoms of the pyridinium ring the hydrogen atoms bonded to the first carbon of the butyl chain are responsible for a strong cation-anion pairing. We believe that pairing energy can be reduced and, consequently, the properties of ILs can be significantly modulated, if the corresponding hydrogen atoms are substituted by fluorine atoms.

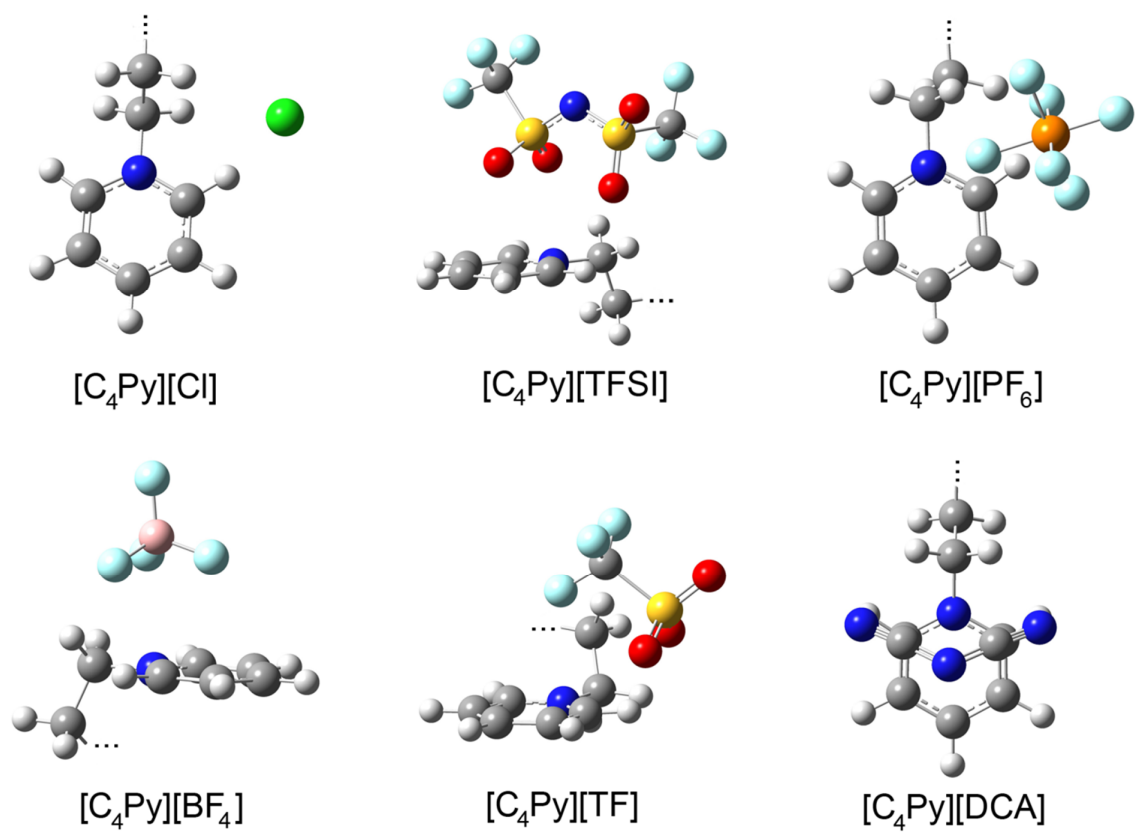

Figure 3. The configurations of ion pairs, corresponding to the internal energy minimum in the absence of thermal motion.

**Performance of the refined force field**

Tables 3 and 4 summarize the thermodynamic (density, heat of vaporization) and transport

(diffusion, viscosity, conductivity) properties obtained using the refined models of each IL. Table 5 lists the experimental data from all sources, which we were able to locate. The force field devised in this work represents a systematic improvement over the original CL&P FF. Although not all experimental values are perfectly reproduced, drastically underestimated ionic transport has been completely eliminated as a result of the correctly captured cation-anion binding. A significant discrepancy with the experimental value was found for self-diffusion coefficients of [$C_4$PY][TFSI] at 353 K. However, temperature dependence of the simulated shear viscosity of this same IL is nearly perfect. Refined non-bonded interactions, such as hydrogen-fluorine, allowed to attain a good to excellent coincidence with the available density data. The observed deviations in any property may be addressed on one-to-one basis, provided that experimental results are trustworthy.

Table 3. Transport properties of the pyridinium-based ionic liquids derived using the refined force field

| IL | | 298 | 313 | 338 | 353 | 403 |
|---|---|---|---|---|---|---|
| | | \multicolumn{5}{c}{T, K} |
| \multicolumn{7}{c}{$\sigma$, S m$^{-1}$} |
| [$C_4$PY][$BF_4$] | | — | 0.43$_5$ | — | 1.2$_1$ | — |
| [$C_4$PY][TFSI] | | — | 0.29$_2$ | — | 0.7$_1$ | — |
| [$C_4$PY][TF] | | 0.23$_1$ | — | 0.7$_1$ | — | — |
| [$C_4$PY][$PF_6$] | | — | 0.12$_1$ | — | 0.54$_3$ | — |
| [$C_4$PY][DCA] | | — | 0.50$_3$ | — | 2.5$_3$ | — |
| [$C_4$PY][Cl] | | — | — | — | 0.04$_1$ | 0.05$_1$ |
| \multicolumn{7}{c}{$\eta$, cP} |
| [$C_4$PY][$BF_4$] | | 93$_7$ | 45$_5$ | 20$_3$ | 11$_2$ | 4.3$_1$ |
| [$C_4$PY][TFSI] | | 54$_6$ | 29$_4$ | 13$_2$ | 9.7$_7$ | 4.0$_6$ |
| [$C_4$PY][TF] | | 43$_6$ | 24$_3$ | 13$_2$ | 9.0$_8$ | 3.6$_6$ |
| [$C_4$PY][$PF_6$] | | 230$_{50}$ | 88$_9$ | 40$_5$ | 25$_7$ | 7.9$_8$ |
| [$C_4$PY][DCA] | | 44$_6$ | 21$_3$ | 10$_2$ | 7.4$_7$ | 2.9$_5$ |
| [$C_4$PY][Cl] | | — | — | — | 710$_{90}$ | 220$_{60}$ |
| \multicolumn{7}{c}{$D_\pm$, 10$^{11}$ m$^2$s$^{-1}$} |
| [$C_4$PY][$BF_4$] | + | — | 1.7$_2$ | — | 7.4$_3$ | — |
| | − | — | 1.2$_2$ | — | 6.0$_7$ | — |
| [$C_4$PY][TFSI] | + | — | 1.73$_6$ | — | 6.7$_3$ | — |
| | − | — | 1.2$_1$ | — | 5.1$_3$ | — |
| [$C_4$PY][TF] | + | 1.22$_4$ | — | 6.0$_2$ | — | — |
| | − | 0.79$_1$ | — | 4.4$_1$ | — | — |
| [$C_4$PY][$PF_6$] | + | — | 0.34$_4$ | — | 2.6$_3$ | — |
| | − | — | 0.19$_3$ | — | 2.0$_2$ | — |

| [C4PY][DCA] | + | — | 3.4$_2$ | — | 14.8$_7$ | — |
|---|---|---|---|---|---|---|
| | − | — | 4.1$_1$ | — | 17.9$_6$ | — |
| [C4PY][Cl] | + | — | — | — | 0.07$_1$ | 0.25$_1$ |
| | − | — | — | — | 0.05$_1$ | 0.23$_1$ |

Since shear viscosity, self-diffusion, and ionic conductivity are largely proportional to each other, we can compare ILs using viscosities over a range of commonly used temperatures (298-403 K). Viscosities of [C4PY][Cl] at 298-338 K are omitted, because these temperatures are outside liquid state range. High viscosity, in essence, is an undesired feature for most applications of ILs. It decreases in the following raw of anions: [Cl]>[PF$_6$]>[BF$_4$]>[TFSI]>[TF]>[DCA]. The first in the list are inorganic anions with strongly polar bonds and charge separation. These anions tend to create hydrogen bonds with the pyridinium ring. They exhibit well-defined atom-atom contacts, which foster crystal-like ordering, decrease ionic transport and increase freezing/glass transition points. In turn, the anions at the end of the list feature more delocalized electron densities, resulting in a softer structure and lower viscosity. The removal/blocking of the strong atom-to-atom contacts is, perhaps, the main instrument in engineering new ILs with ever wider liquid temperature range. [C4PY][DCA] is an interesting compound, thanks to its modest shear viscosity and very high conductivity (2.5 S m$^{-1}$ at 353 K). Experimental investigation of this IL is deemed desirable.

Table 4. Summary of thermodynamic properties derived using the new force field

| T, K | [C4PY][BF$_4$] | [C4PY][TFSI] | [C4PY][TF] | [C4PY][PF$_6$] | [C4PY][DCA] | [C4PY][Cl] |
|---|---|---|---|---|---|---|
| *d*, kg m$^{-3}$ | | | | | | |
| 313 | 1185 | 1457 | 1307 | 1377 | 1018 | 1095 |
| 353 | 1151 | 1415 | 1269 | 1337 | 987 | 1077 |
| ΔH$_{vap}$, kJ mol$^{-1}$ | | | | | | |
| 313 | 139 | 140 | 130 | 151 | 132 | 172 |

Table 5. Transport and thermodynamics properties of ILs at various temperatures derived using direct physical experiments

| T, K | *d*, kg·m$^{-3}$ | ΔH$_{vap}$, kJ·mol$^{-1}$ | $D_+$, 10$^{-11}$, m$^2$·s$^{-1}$ | $D_-$, 10$^{-11}$, m$^2$·s$^{-1}$ | $\eta$, cP | $\sigma$, S·m$^{-1}$ |
|---|---|---|---|---|---|---|
| \[C4PY\]\[Cl\] | | | | | | |
| 298 | | 169.7 [25] | | | | |
| \[C4PY\]\[BF4\] | | | | | | |
| 298 | 1213.4 [26]<br>1214.4 [27]<br>1220 [28]<br>1222 [29]<br>1214 [30] | 167 [31] | 0.9 [28] | 1.0 [28] | 163.28 [26]<br>102.7 [28]<br>161.6 [30] | 0.191 [28]<br>0.241 [32] |
| 313 | 1203.2 [26]<br>1205.3 [27]<br>1203 [33]<br>1208 [28]<br>1203.7 [34]<br>1204 [30] | | 1.9 [28] | 2.1 [28] | 70.29 [26]<br>51.6 [28]<br>71.11 [34]<br>71.7 [30] | 0.402 [28]<br>0.516 [32] |
| 338 | 1186.3 [26]<br>1188 [28]<br>1187 [30] | | 5.1 [28] | 5.8 [28] | 26.01 [26]<br>21.9 [28]<br>26.1 [30] | 0.995 [28]<br>1.306 [32] |
| 353 | 1176.0 [26]<br>1176 [28]<br>1176 [30] | | 8.3 [28] | 9.4 [28] | 15.84 [26]<br>14.8 [28]<br>16.4 [30] | 1.500 [28]<br>1.989 [32] |
| \[C4PY\]\[TFSI\] | | | | | | |
| 298 | 1449 [28]<br>1454 [35] | 154* [31]<br>138.1 [36] | 2.4 [28]<br>2.2 [35] | 2.0 [28]<br>1.8 [35] | 56.8 [28]<br>59.0 [35] | 0.229 [28]<br>0.320 [35] |
| 313 | 1436 [28]<br>1440 [35] | | 4.3 [28]<br>4.1 [35] | 3.7 [28]<br>3.3 [35] | 31.0 [28]<br>31.4 [35] | 0.398 [28]<br>0.560 [35] |
| 338 | 1414 [28]<br>1417 [35] | | 9.7 [28]<br>9.2 [35] | 8.6 [28]<br>7.6 [35] | 14.2 [28]<br>14.3 [35] | 0.769 [28]<br>1.13 [35] |
| 353 | 1401 [28]<br>1403 [35] | 131.1 [399.5 36] | 14.3 [28]<br>14 [35] | 13.0 [28]<br>11 [35] | 9.8 [28]<br>9.9 [35] | 1.03 [28]<br>1.57 [35] |
| \[C4PY\]\[TF\] | | | | | | |
| 298 | 1314 [30] | | | | 127.9 [30] | 0.207 [32] |
| 313 | 1303 [30] | | | | 61.9 [30] | 0.409 [32] |
| 338 | 1283 [30] | | | | 24.7 [30] | 0.970 [32] |
| 353 | 1272 [30] | | | | 16.1 [30] | 1.45 [32] |

* The value was predicted using a theory, in which $\Delta H_{vap}$ is represented by a sum of Coulombic contribution and two van der Waals contributions (from cation and from anion).

Figure 4 depicts selected radial distribution functions (RDFs) computed for the cation-anion coordination sites. The primary goal of these RDFs is to show that ion-ion coordination implemented by the newly proposed force field is in an agreement with the DFT-optimized geometries (Figure 3). The heights of peaks for the hydrogen bonding atom pairs are probably underestimated, because the combination of only Coulomb and Lennard-Jones pairwise terms cannot in full reproduce a potential energy surface of a hydrogen bond. This is particularly true for the cases of the strong and short hydrogen bonds, where an admixture of a covalent bonding must be anticipated. The classical descriptions miss a covalent bond term. Electronic structure based molecular dynamics simulations may provide a valuable additional knowledge of this phenomenon. The underestimated height is compensated by an increased peak width, so that the first shell coordination numbers do not change. A remarkable feature of all ILs is that, unlike in conventional salts (melts) and inorganic electrolyte solutions, a coordination shell of an ion cannot be clearly defined. It is demonstrated by the RDFs, none of which decays to zero beyond the first maximum. Therefore, ions belonging to the first coordination shell cannot be univocally separated from the ions of the second coordination shell.

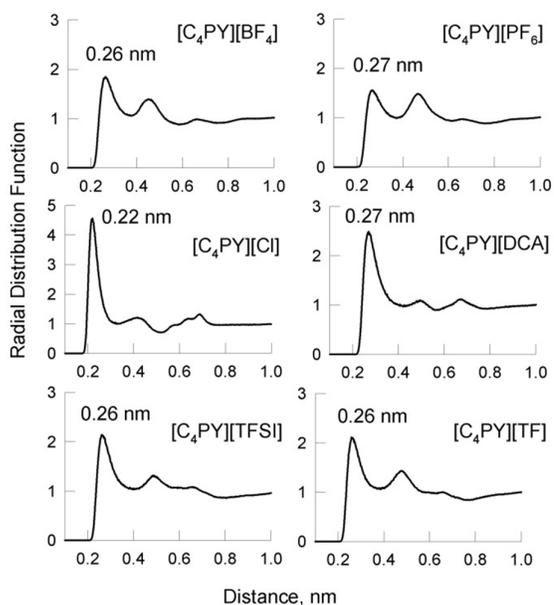

Figure 4. Radial distributed functions computed for the most positively charged atom of the cation and the most negatively charged atom of the anion: H-F in [C$_4$PY][BF4], H-F in [C$_4$PY][PF$_6$], H-Cl in [C$_4$PY][Cl], H-N in [C$_4$PY][DCA], H-O in [C$_4$PY][TFSI], and H-O in [C$_4$PY][TF]. All RDFs were computed at 298 K, except for [C$_4$PY][Cl], which was computed at 353 K.

**Conclusions**

We started from the Lopes—Padua force field for pyridinium-based ionic liquids and elaborated a procedure to account for specific cation-anion interactions in the liquid state of each

IL. The introduced refinement allowed for good reproduction of key transport properties, such as self-diffusion coefficients, shear viscosity, and ionic conductivity. Thermodynamics (density and heat of vaporization) and structure (especially, halogen-hydrogen contact distance) have been additionally improved in a few cases. This work suggests that cation-anion interactions must be refined for each IL separately. In turn, parameters, derived for isolated cation and isolated anion, provide a less accurate description. The force field is obtained using the original Lopes—Padua parameters and hybrid density functional theory calculations of the IL ion pairs. No experimental data have been used during the FF development to guarantee consistency and physical soundness. The developed force field can be applied to a wide variety of computer simulations. Furthermore, the models are fully compatible with a large set of cations and anions available in the CL&P FF.

Note, there are several procedures to fit ESP using point charges. Different procedures provide somewhat different charge distributions. In turn, different electronic structure methods coupled with different basis sets impact the ESP to be fitted. One should not expect that the scaling factors derived in the present work are absolute quantities. Additionally, point ESP charges and charge transfer effects are conformation dependent. Only a single local-minimum conformation of each IL was considered for the charge refinement in our procedure.

Extension of the proposed methodology to other families of ionic liquids is underway.

**Acknowledgments**

Iu.V.V. acknowledges a research grant from FCT with the reference SFRH/BPD/97918/2013. MEMPHYS is the Danish National Center of Excellence for Biomembrane Physics. The Center is supported by the Danish National Research Foundation. The Ukrainian-American Laboratory of Computational Chemistry (UALCC) is acknowledged for providing a computational platform.

**Supporting Information**

The force field derived in this work and ready-to-use topology input files for the GROMACS molecular dynamics simulation package are available free of charge upon e-mail to vvchaban@gmail.com. Please, include your name and academic affiliation in the message.

Figures S1-S3 referenced in the manuscript can be found in Supporting Information. This information is available free of charge via the Internet at http://pubs.acs.org.

**Contact Information**

E-mail for correspondence: vvchaban@gmail.com; chaban@sdu.dk (V.V.C.).

# REFERENCES

(1) Chaban, V. V.; Prezhdo, O. V. Ionic and Molecular Liquids: Working Together for Robust Engineering. *Journal of Physical Chemistry Letters* **2013**, *4*, 1423-1431.
(2) Chaban, V. Polarizability Versus Mobility: Atomistic Force Field for Ionic Liquids. *Physical Chemistry Chemical Physics* **2011**, *13*, 16055-16062.
(3) Lopes, J. N. C.; Padua, A. A. H. Cl&P: A Generic and Systematic Force Field for Ionic Liquids Modeling. *Theoretical Chemistry Accounts* **2012**, *131*.
(4) Maciel, C.; Fileti, E. E. Molecular Interactions between Fullerene C-60 and Ionic Liquids. *Chemical Physics Letters* **2013**, *568*, 75-79.
(5) Lopes, J. N. C.; Padua, A. A. H. Molecular Force Field for Ionic Liquids Composed of Triflate or Bistriflylimide Anions. *Journal of Physical Chemistry B* **2004**, *108*, 16893-16898.
(6) Hantal, G.; Voroshylova, I.; Cordeiro, M. N. D. S.; Jorge, M. A Systematic Molecular Simulation Study of Ionic Liquid Surfaces Using Intrinsic Analysis Methods. *Physical Chemistry Chemical Physics* **2012**, *14*, 5200-5213.
(7) Lopes, J. N. C.; Padua, A. A. H. Molecular Force Field for Ionic Liquids Iii: Imidazolium, Pyridinium, and Phosphonium Cations; Chloride, Bromide, and Dicyanamide Anions. *Journal of Physical Chemistry B* **2006**, *110*, 19586-19592.
(8) Darvas, M.; Jorge, M.; Cordeiro, M. N. D. S.; Kantorovich, S. S.; Sega, M.; Jedlovszky, P. Calculation of the Intrinsic Solvation Free Energy Profile of an Ionic Penetrant across a Liquid-Liquid Interface with Computer Simulations. *Journal of Physical Chemistry B* **2013**, *117*, 16148-16156.
(9) Lopes, J. N. C.; Deschamps, J.; Padua, A. A. H. Modeling Ionic Liquids of the 1-Alkyl-3-Methylimidazolium Family Using an All-Atom Force Field. *Ionic Liquids Iiia: Fundamentals, Progress, Challenges, and Opportunities, Properties and Structure* **2005**, *901*, 134-149.
(10) Lopes, J. N. C.; Deschamps, J.; Padua, A. A. H. Modeling Ionic Liquids Using a Systematic All-Atom Force Field. *Journal of Physical Chemistry B* **2004**, *108*, 2038-2047.
(11) Lopes, J. N. C.; Deschamps, J.; Padua, A. A. H. Modeling Ionic Liquids Using a Systematic All-Atom Force Field (Vol 104b, Pg 2038, 2004). *Journal of Physical Chemistry B* **2004**, *108*, 11250-11250.
(12) Lopes, J. N. C.; Padua, A. A. H.; Shimizu, K. Molecular Force Field for Ionic Liquids Iv: Trialkylimidazolium and Alkoxycarbonyl-Imidazolium Cations; Alkylsulfonate and Alkylsulfate Anions. *Journal of Physical Chemistry B* **2008**, *112*, 5039-5046.
(13) Shimizu, K.; Almantariotis, D.; Gomes, M. F. C.; Padua, A. A. H.; Lopes, J. N. C. Molecular Force Field for Ionic Liquids V: Hydroxyethylimidazolium, Dimethoxy-2-Methylimidazolium, and Fluoroalkylimidazolium Cations and Bis(Fluorosulfonyl)Amide, Perfluoroalkanesulfonylamide, and Fluoroalkylfluorophosphate Anions. *Journal of Physical Chemistry B* **2010**, *114*, 3592-3600.
(14) Sambasivarao, S. V.; Acevedo, O. Development of Opls-Aa Force Field Parameters for 68 Unique Ionic Liquids. *Journal of Chemical Theory and Computation* **2009**, *5*, 1038-1050.
(15) Borodin, O. Polarizable Force Field Development and Molecular Dynamics Simulations of Ionic Liquids. *Journal of Physical Chemistry B* **2009**, *113*, 11463-11478.
(16) Chaban, V. V.; Voroshylova, I. V.; Kalugin, O. N. A New Force Field Model for the Simulation of Transport Properties of Imidazolium-Based Ionic Liquids. *Physical Chemistry Chemical Physics* **2011**, *13*, 7910-7920.
(17) Chaban, V. V.; Prezhdo, O. V. A New Force Field Model of 1-Butyl-3-Methylimidazolium Tetrafluoroborate Ionic Liquid and Acetonitrile Mixtures. *Physical Chemistry Chemical Physics* **2011**, *13*, 19345-19354.
(18) Chaban, V. V.; Voroshyloya, I. V.; Kalugin, O. N.; Prezhdo, O. V. Acetonitrile Boosts Conductivity of Imidazolium Ionic Liquids. *Journal of Physical Chemistry B* **2012**, *116*, 7719-7727.

(19)	Hess, B. Determining the Shear Viscosity of Model Liquids from Molecular Dynamics Simulations. *Journal of Chemical Physics* **2002**, *116*, 209-217.
(20)	Bussi, G.; Donadio, D.; Parrinello, M. Canonical Sampling through Velocity Rescaling. *Journal of Chemical Physics* **2007**, *126*, 014101.
(21)	Parrinello, M.; Rahman, A. Polymorphic Transitions in Single Crystals: A New Molecular Dynamics Method. *J. Appl. Phys.* **1981**, *52*.
(22)	Hess, B.; Kutzner, C.; van der Spoel, D.; Lindahl, E. Gromacs 4: Algorithms for Highly Efficient, Load-Balanced, and Scalable Molecular Simulation. *Journal of Chemical Theory and Computation* **2008**, *4*, 435-447.
(23)	Chai, J. D.; Head-Gordon, M. Long-Range Corrected Double-Hybrid Density Functionals. *Journal of Chemical Physics* **2009**, *131*, 174105.
(24)	Breneman, C. M.; Wiberg, K. B. Determining Atom-Centered Monopoles from Molecular Electrostatic Potentials - the Need for High Sampling Density in Formamide Conformational-Analysis. *Journal of Computational Chemistry* **1990**, *11*, 361-373.
(25)	Verevkin, S. P.; Ralys, R. V.; Emel'yanenko, V. N.; Zaitsau, D. H.; Schick, C. Thermochemistry of the Pyridinium- and Pyrrolidinium-Based Ionic Liquids. *Journal of Thermal Analysis and Calorimetry* **2013**, *112*, 353-358.
(26)	Mokhtarani, B.; Sharifi, A.; Mortaheb, H. R.; Mirzaei, M.; Mafi, M.; Sadeghian, F. Density and Viscosity of Pyridinium-Based Ionic Liquids and Their Binary Mixtures with Water at Several Temperatures. *Journal of Chemical Thermodynamics* **2009**, *41*, 323-329.
(27)	Gu, Z.; Brennecke, J. F. Volume Expansivities and Isothermal Compressibilities of Imidazolium and Pyridinium-Based Ionic Liquids. *Journal of Chemical & Engineering Data* **2002**, *47*, 339-345.
(28)	Noda, A.; Hayamizu, K.; Watanabe, M. Pulsed-Gradient Spin-Echo 1h and 19f Nmr Ionic Diffusion Coefficient, Viscosity, and Ionic Conductivity of Non-Chloroaluminate Room-Temperature Ionic Liquids. *Journal Physical Chemistry B* **2001**, *105*, 4603-4610.
(29)	Ye, C. F.; Shreeve, J. M. Rapid and Accurate Estimation of Densities of Room-Temperature Ionic Liquids and Salts. *Journal of Physical Chemistry A* **2007**, *111*, 1456-1461.
(30)	Bandrés, I.; Royo, F. l. M.; Gascón, I.; Castro, M.; Lafuente, C. Anion Influence on Thermophysical Properties of Ionic Liquids: 1-Butylpyridinium Tetrafluoroborate and 1-Butylpyridinium Triflate. *The Journal of Physical Chemistry B* **2010**, *114*, 3601-3607.
(31)	Deyko, A.; Lovelock, K. R. J.; Corfield, J. A.; Taylor, A. W.; Gooden, P. N.; Villar-Garcia, I. J.; Licence, P.; Jones, R. G.; Krasovskiy, V. G.; Chernikova, E. A. et al. Measuring and Predicting Delta H-Vap(298) Values of Ionic Liquids. *Physical Chemistry Chemical Physics* **2009**, *11*, 8544-8555.
(32)	Bandrés, I.; Montaño, D. F.; Gascón, I.; Cea, P.; Lafuente, C. Study of the Conductivity Behavior of Pyridinium-Based Ionic Liquids. *Electrochimica Acta* **2010**, *55*, 2252-2257.
(33)	Blanchard, L. A.; Gu, Z.; Brennecke, J. F. High-Pressure Phase Behavior of Ionic Liquid/Co2 Systems. *The Journal of Physical Chemistry B* **2001**, *105*, 2437-2444.
(34)	García-Mardones, M.; Gascón, I.; López, M. C.; Royo, F. M.; Lafuente, C. Viscosimetric Study of Binary Mixtures Containing Pyridinium-Based Ionic Liquids and Alkanols. *Journal of Chemical & Engineering Data* **2012**, *57*, 3549-3556.
(35)	Tokuda, H.; Ishii, K.; Susan, M. A. B. H.; Tsuzuki, S.; Hayamizu, K.; Watanabe, M. Physicochemical Properties and Structures of Room-Temperature Ionic Liquids. 3. Variation of Cationic Structures. *The Journal of Physical Chemistry B* **2006**, *110*, 2833-2839.
(36)	Zaitsau, D. H.; Yermalayeu, A. V.; Emel'yanenko, V. N.; Verevkin, S. P.; Welz-Biermann, U.; Schubert, T. Structure-Property Relationships in Ils: A Study of the Alkyl Chain Length Dependence in Vaporisation Enthalpies of Pyridinium Based Ionic Liquids. *Science China-Chemistry* **2012**, *55*, 1525-1531.

TOC Image

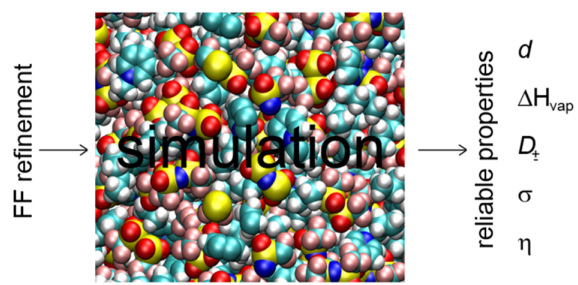